\pdfoutput=1
\documentclass[reprint,superscriptaddress,noeprint]{revtex4-1}
\usepackage[utf8]{inputenc}
\usepackage[dvipsnames]{xcolor}
\usepackage{dcolumn}
\usepackage{bm}
\usepackage[final]{graphicx}
\usepackage{times,bbm,amsmath,amssymb}
\usepackage{epsfig,color}
\usepackage{float,siunitx}
\usepackage[caption = false]{subfig}
\usepackage[english]{babel}
\usepackage{thumbpdf,enumerate}
\usepackage{booktabs}
\usepackage{sidecap}
\usepackage[scaled=.8]{couriers}
\usepackage{pstricks}
\usepackage{multirow}
\usepackage{placeins}
\usepackage{pst-grad}
\usepackage{epigraph}
\usepackage{longtable}
\usepackage{booktabs}
\usepackage{multirow}
\usepackage{gensymb}
\usepackage{array}

\usepackage[colorlinks=true,bookmarks=false,linkcolor=RoyalBlue,urlcolor=RoyalBlue,citecolor=RoyalBlue,breaklinks]{hyperref}

\begin{document}

\newcommand{\FA}[1]{\textcolor{teal}{#1}}

\title{Experimental quantum-enhanced response function estimation}

\author{I. Gianani}
\affiliation{Dipartimento di Scienze, Universit\'a degli Studi Roma Tre, Via della Vasca Navale 84, 00146, Rome, Italy}
\email{ilaria.gianani@uniroma3.it}
\author{F. Albarelli}
\affiliation{Faculty of Physics, University of Warsaw, Pasteura 5, PL-02-093 Warszawa, Poland}
\author{V. Cimini}
\affiliation{Dipartimento di Scienze, Universit\'a degli Studi Roma Tre, Via della Vasca Navale 84, 00146, Rome, Italy}
\author{M. Barbieri}
\affiliation{Dipartimento di Scienze, Universit\'a degli Studi Roma Tre, Via della Vasca Navale 84, 00146, Rome, Italy}
\affiliation{Istituto Nazionale di Ottica - CNR, Largo Enrico Fermi 6, 50125 Florence, Italy}

\maketitle

\textbf{ Characterizing a system often demands learning its response function to an applied field. Such knowledge is rooted on the experimental evaluation of punctual fiducial response and interpolation to access prediction at arbitrary values. Quantum metrological resources are known to provide enhancement in assessing these fiducial points~\cite{Giovannetti1,Giovannetti3,paris,lee}, but the implications for improved function estimation have only recently been explored~\cite{ueda,Tsang2011}, and have not been yet demonstrated. Here we show an experimental realization of function estimation based on a photonic achitecture. The phase response of a liquid-crystal to a voltage has been reconstructed by means of quantum and classical phase estimation, providing evidence of the superiority of the former and highlighting the interplay between punctual statistical error and interpolation error. Our results show how quantum resources should successfully be employed to access the rich information contained in continuous signals.}

In the quest for superior quantum technology, the development of sensors showing a palpable advantage has reached the state of solid demonstrations.
The basic operations allowing for quantum metrological enhancement have indisputably been validated in photonics~\cite{Polinoreview,multireview,pirandola,Demkowicz-Dobrzanski2015a} and field sensing~\cite{atomicensembles,qsensingreview}. Current development is aiming, on the one hand, at consolidating the technological readiness, on the other at exploring new paradigms, based on the results attained so far. This is akin to the workflow in computing: once elementary operations are mastered, these are then combined to deliver more elaborated capabilities.

In quantum photonics, phase estimation can indeed serve as such basic element. Combining several phase estimation routines gives way to addressing novel problems, ranging from multiple phase estimation in interferometers~\cite{peter}, phase tracking~\cite{Wheatley2010,Yonezawa2012,Cimini20191}, and, as recently introduced, function estimation~~\cite{ueda}. This latter 
opens up unexplored opportunities for quantum enhancement in important problems such as evaluating time-dependent signals and mapping fields~\cite{Tsang2011,Petersen2006,Berry2013a,Berry2015a,Ng2016,MartinCiurana2017,Laverick2018}. 

In this Letter, we realise quantum function estimation, showing that quantum advantage is attained only if resources are cleverly distributed. An unconditional advantage could be obtained within the current technological effort~\cite{sergei}, provided that sensors are used accounting for both the uncertainty on measured points, and the issue of interpolating between them. This work lays the ground for the inclusion of quantum estimation in functional data analysis~\cite{libro} in the near future.

\begin{figure*}[t!]
  \includegraphics[width=\textwidth]{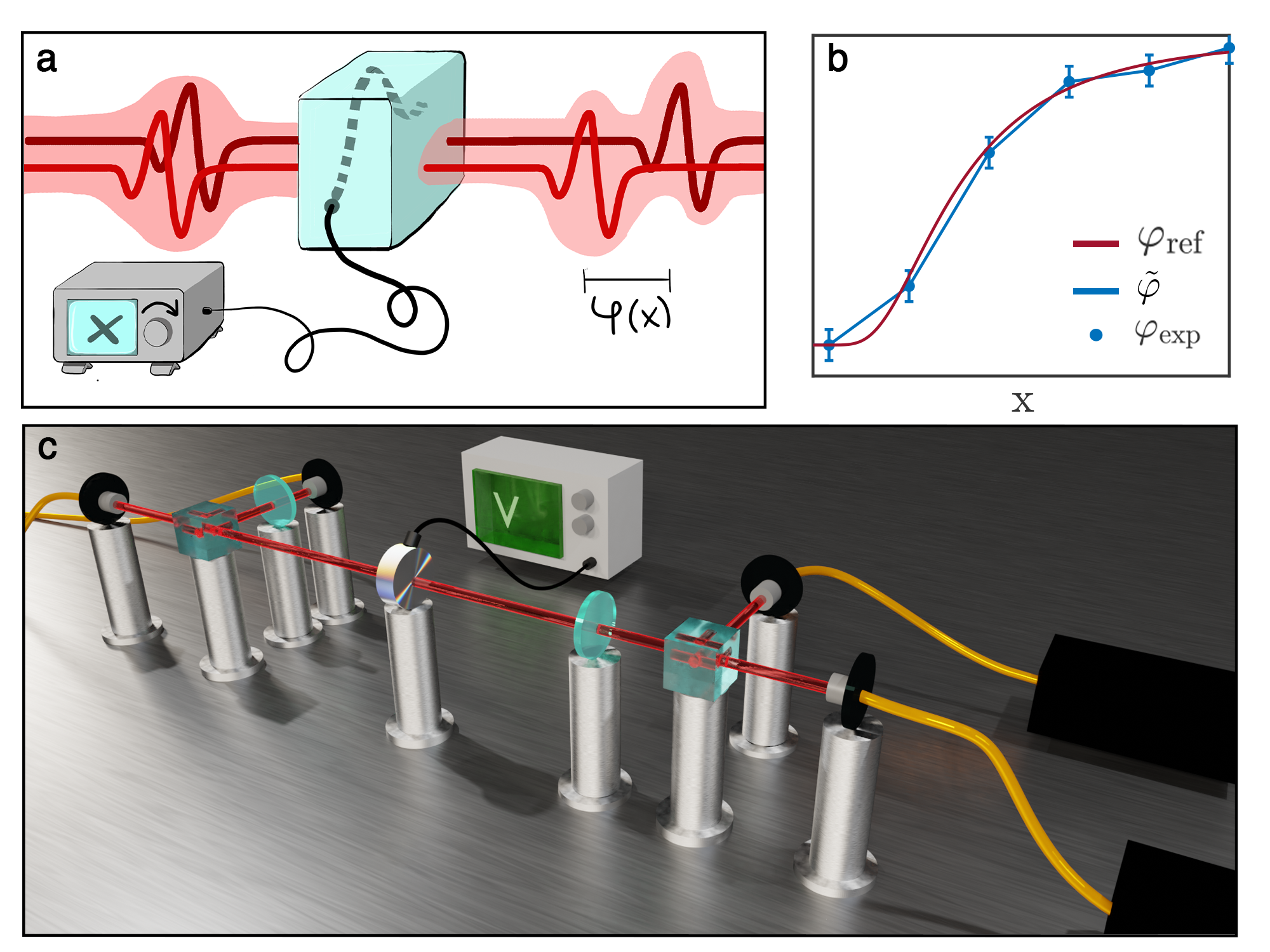}
    \caption{{\bf Function estimation.} {\bf a.} {\it Schematic representation.} A system's response function to an applied voltage, $\varphi(x)$, is estimated as a relative phase with probe states. {\bf b.} {\it Function estimation.} The aim of the experiment is to produce an estimator for the function $\varphi(x)$ based on a set of fiducial points to which a statistical error is associated due to the limited amount of resources-per-point adopted.{\bf c.} {\it Experimental Setup.}}
    \label{fig:fig1}
\end{figure*}

Function estimation  is formally a generalization of multiparameter metrology~\cite{Szczykulska2016,multireview,Demkowicz-Dobrzanski2020} and it has been intensively studied in the context of time-dependent signals. Here, we follow the analysis of~\cite{ueda}, which is not specific to time-dependent signals. Consider a system whose response function to an applied signal $x$ is indeed a phase $\varphi(x)$ (Fig.~\ref{fig:fig1}a): if enough experimental values $\varphi_{\rm exp}(x)$ are collected as fiducial references, one can interpolate them in order to obtain an estimate $\tilde \varphi(x)$ of the function for any values of the signal in a given range $[ 0,L ]$ (Fig.~\ref{fig:fig1}b).
The accuracy on $\tilde \varphi(x)$ will depend on the interpolation method, the number of sampling points of the signal, and on the uncertainty on the measured values. Thus, when a fixed number of resources are allocated, it should be ensured that they are optimally deployed taking into account these contrasting error sources.

For any given value of $x$, quantum phase estimation guarantees that an advantage can be attained \textit{punctually} on the system response.
By leveraging on these ameliorated performance, the estimation of the whole function can also be improved.
The most straightforward implication is that, for fixed resources, the statistical error on the reference points will decrease when using quantum light.
There is however a more involved effect, originating in the interplay between the distinct sources of error: for a given amount of resources, and given acceptable statistical uncertainty on the fiducial points, quantum light allows to increase the sampling density.
In summary, quantum resources can provide either better punctual estimations on a given number of sampled points, or more sampled points with the same uncertainty, with respect their classical penchant.



A scenario for which this approach is relevant  is described in Fig.~\ref{fig:fig1}c. We consider the response function of a liquid crystal device to an applied voltage which can be controlled and swept across the range $[0, 3]\,V$ as the birefringent phase associated to its optical axes.
We set the crystal so that the voltage-induced fast and slow axis are oriented along the diagonal (D) and antidiagonal (A) polarizations.
This phase is estimated based on measurements on individual photons, providing the classical limit, and N00N states, exhibiting quantum advantage.
These are generated with a noncollinear type-I spontaneous parametric downconversion (SPDC) source. Individual photons are heralded on one of the modes of the SPDC, and are prepared in the horizontal polarization, i.e. the superposition of D and A:  $\vert \psi_1 \rangle = \left(\vert 1 \rangle_D\vert 0 \rangle_A + \vert 0 \rangle_D \vert 1 \rangle_A\right)/\sqrt{2}.$ In order to produce the N00N states, we make use of both photons from SPDC which are superposed with orthogonal polarizations on a polarizing beam splitter (PBS). This operation prepares them in a N00N state in the diagonal basis: $\vert \psi_2 \rangle = \left(\vert 2 \rangle_D\vert 0 \rangle_A + \vert 0 \rangle_D \vert 2 \rangle_A\right)/\sqrt{2}.$ The relative phase is accumulated twice as fast in the state $\vert \psi_2 \rangle$ than in $\vert \psi_1 \rangle$, resulting in superior sensitivity.

\begin{figure*}[t!]
    \includegraphics[width=\textwidth]{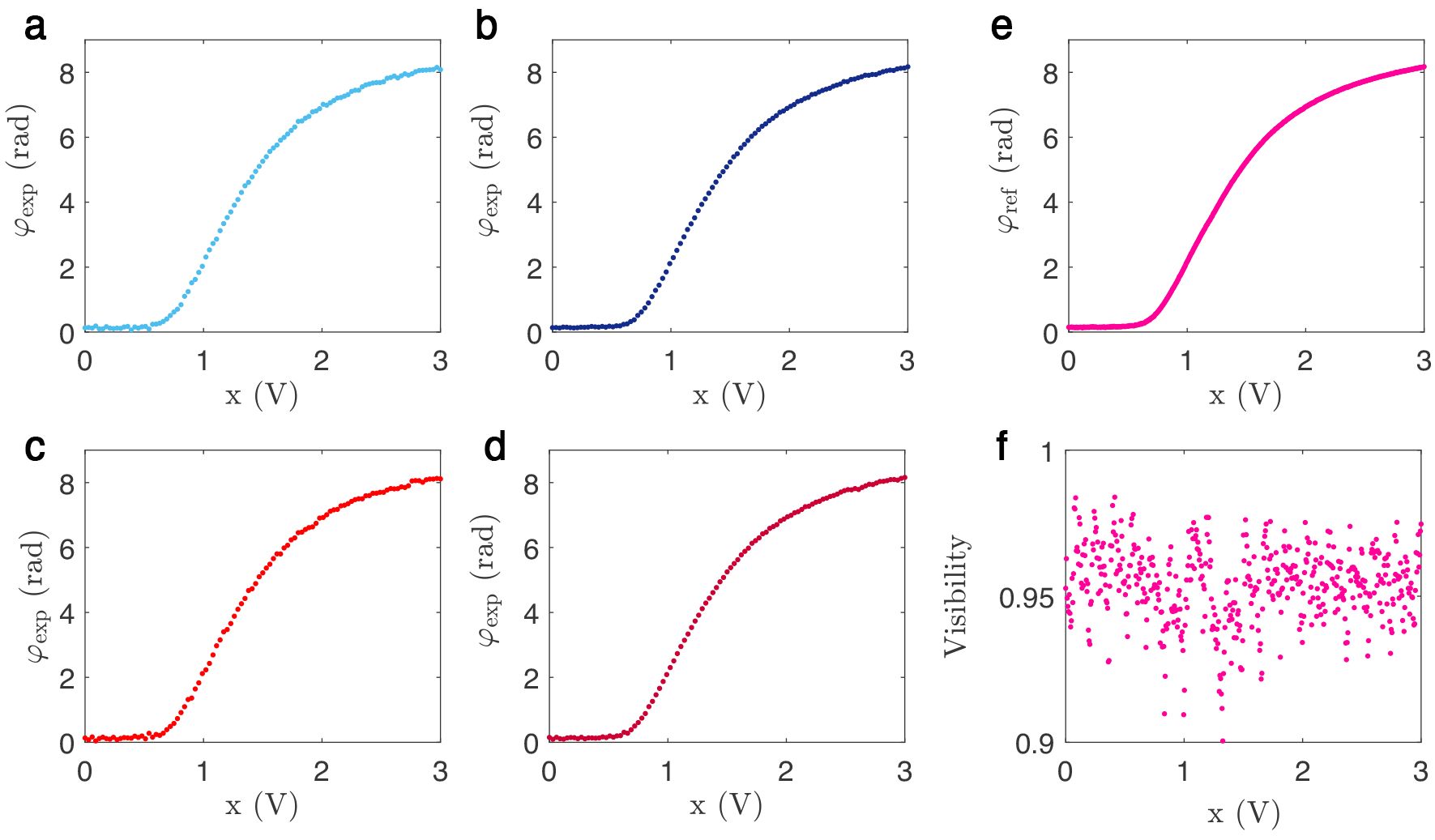}
    \caption{{\bf Bayesian phase estimation.} Measured birefringent phases imparted by the liquid crystal device at different voltage values, estimated via a multiparameter Bayesian approach, as detailed in the Methods section, for: \textbf{a.} N00N states with $N_r=800$, $\Delta x =0.03\,V$ {\bf b.} N00N states with $N_r=1900$, , $\Delta x =0.03\,V${\bf c.} individual photons with $N_r=800$, , $\Delta x =0.03\,V$ {\bf d.} individual photons with $N_r=1900$. , $\Delta x =0.03\,V$ \textbf{e.} Measured $\varphi_{\rm ref}(x)$ with N00N states, with $N_s=500$ ($\Delta x =0.006\,V$), and $N_r\simeq 60k$. \textbf{f.} Estimated fringe visibility for the reference measurement in \textbf{e}, which, due to the high statistics and number of sampling point, corresponds to the longest measurement acquired. In all the graphs the errors are smaller than the datapoints.}
    \label{fig:fig2}
\end{figure*}

For each voltage setting $x_i$, the corresponding value of the phase $\phi_i$ is retrieved by means of either $\vert \psi_1 \rangle$ or $\vert \psi_2 \rangle$ through a multiparameter Bayesian routine, which includes the measurement of the fringe visibility~\cite{Roccia2018}: this guarantees to perform an unbiased estimation against the instabilities of the system which might become significant due to the amount of time necessary to accumulate the signal and sampling required.
We acquired 100 points using the same number of resources $N_{r}=800$ and $N_{r}=1900$ both for N00N and single-photon states; this number is given by the number of photons in the state times the number of repetitions of the measurement.
The Bayesian estimated phases from these measurements are presented in Fig.~\ref{fig:fig2}a-b for N00N states, and Fig.~\ref{fig:fig2}d-e for single photons. 

The collected phase values are then employed to obtain estimates of the function $\tilde \varphi(x)$ for arbitrary values, using two different strategies, viz. linear  and nearest-neighbour interpolation. The associated error is then quantified by:
\begin{equation}
    \delta_0^2=\frac{1}{L}\int\, \vert \varphi(x)-\tilde\varphi(x) \vert^2\,dx,
    \label{eq:errore0}
\end{equation}
which is the average quadratic deviation over the whole variable range.
As a matter of fact, the function $\varphi(x)$ is unknown, making the error \eqref{eq:errore0} experimentally inaccessible.
A measurable proxy can be obtained by assessing values of $\varphi(x)$ at much denser sampling and much lower statistical uncertainty than the points used in the experiment: these are identifiable, for all practical purposes, with the true values of the function.
We measure the reference phase $\varphi_{\rm ref}(x)$, using N00N states, acquiring $N_s^{\rm ref}=500$ sampling points adopting $N_r \simeq 60k$ resources for each fiducial point sampled.
The corresponding phase and visibility measurements are shown in Fig.~\ref{fig:fig2}e-f, demonstrating the appropriateness the multiparameter approach.

Consequently, the error in \eqref{eq:errore0} can be approximated by a sum over a discrete set of values of $x$, dictated by the sampling of the reference $\varphi_{\rm ref}(x)$: 
\begin{equation}
\delta^2 = \frac{1}{L}\sum_{x=0}^L  \vert \tilde{\varphi}(x) -\varphi_{\rm ref}(x)\vert^2 \Delta x_{\rm ref}
\label{eq:errore1}
\end{equation}
where $\Delta x_{\rm ref}$ is the sampling resolution of the reference. This is the figure of merit we explored in our experiment.
Differently from the original proposal~\cite{ueda}, we do not focus on the different scaling of the error with the number of photons: in fact, this is an asymptotic property which we can not capture with our probe states.

\begin{figure*}[t!]
    \includegraphics[width=\textwidth]{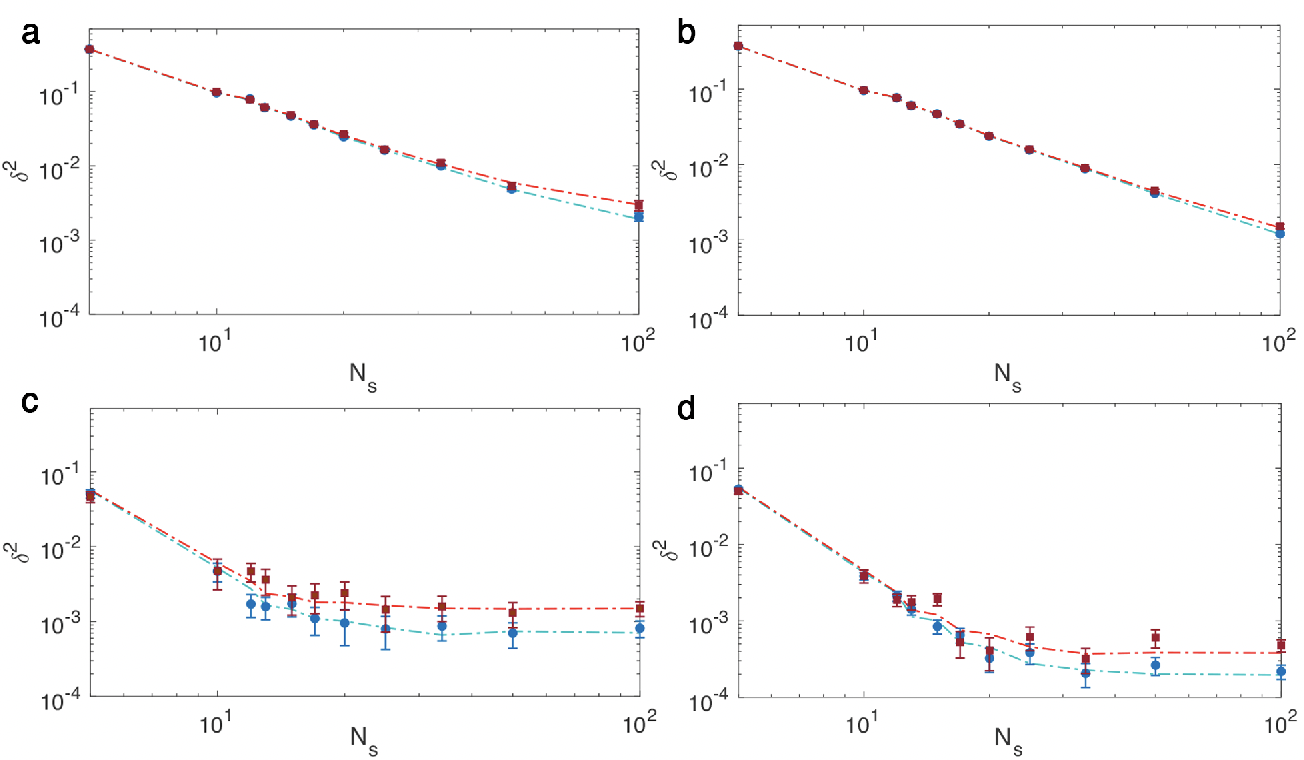}
    \caption{{\bf Function estimation.} Simulated (dotted lines) and measured error $\delta^2$ as a function of the number of sampled points $N_s$ with N00N states (light blue) and individual photons (red) probes for {\bf a,}  $N_r=800$, for nearest-neighbour interpolation,{\bf b,}  $N_r=1900$, for nearest-neighbour interpolation, {\bf c,}  $N_r=800$, for linear interpolation, and {\bf d,}  $N_r=1900$, for linear interpolation. The errors are compatible with what expected from the simulations performed at the Cramér-Rao bound (see Methods).}
    \label{fig:results}
\end{figure*}

Function estimation is carried out as follows: for each set in Fig.~\ref{fig:fig2} we select datapoints so to obtain subsets comprised of a different value of sampling points $N_s$, up to $N_s=100$.
For each subset we interpolate the points to match the sampling of the reference $N_s^{\rm ref}$ using the two methods introduced above.
This strategy then keeps the statistical uncertainty fixed, and adopts increasing resources solely to vary the density of the sampling.

We first discuss the nearest-neighbour interpolation method, generally employed when no assumptions on the regularity of the function are granted, e.g. in the presence of discontinuities.
The behaviour of the error $\delta^2$ for this instance for both values of $N_r$ is reported in Fig.~\ref{fig:results} a-b.
It is evident how the primary source of error up to $N_s=50$ comes from the interpolation, rather than from the statistical uncertainties.
Therefore, if more resources are available, they would be more conveniently used to increase the sampling rate, rather than improving the significance of the individual points.
However, for denser sampling the advantage of quantum light becomes relevant: resources can be allocated to improve the uncertainties.
This analysis mirrors the one in the proposal~\cite{ueda}, which considered this same interpolation method.
For regular functions, a linear interpolation is more suitable.
As shown in Figs.~\ref{fig:results}c-d, fewer points are needed in order to achieve the same accuracy on the function estimation.
The improvement linked to the use of quantum resources becomes relevant earlier, but eventually saturates, due to the statistical error.

To conclude, we have a presented a proof-of-principle estimation of a simple function based upon photonic metrology.
In this experiment we highlight the crucial interplay between statistical and interpolation errors, which becomes evident already at relatively low sampling densities. Increasing $N_s$ can improve the estimation only up to a limit in which the statistical uncertainty on the individual points becomes the predominant source, as determined by the complexity of the function. 

Quantum advantage is only beneficial once this level is attained. If one is interested only in a few scalar quantities obtained from the unknown function, the most useful approach is semiparametric estimation, recently generalized to the quantum domain~\cite{Tsang2019,Tsang2019b}.

In real-life problems, such as probing of the response of biological systems~\cite{Taylor2016}, more complicated functions will be sampled and the advantage of punctual reduced statistical uncertainty will likely become manifest at higher sampling densities. Therefore, to minimize the overall error~\eqref{eq:errore0} it will be fundamental to take into account the interpolation, as we have shown, to make a judicious choice about the allocation of the resources.









\section*{Methods}

\textbf{Single photon source.} A CW Diode laser at 405 nm pumps a 3-mm $\beta$-barium borate (BBO) crystal cut for noncollinear Type I phasematching, generating via spontaneous parametric down conversion (SPDC) two degenerate photons at 810 nm. The laser power is set to 50 mW for the reference measurement, and to 6 mW and 13 mW respectively for the  measurements at $N_r=800$ an $N_r=1900$. The two modes are then selected through interference filters with FWHM=7.3 nm and single mode fibers. \\

\textbf{Interferometer.} For the N00N measurements, a polarization interferometer is setup as shown in Fig.~\ref{fig:fig1} c. of the main text: the polarization of one mode is rotated by means of a half-wave plate (HWP) and the two photons are then overlapped on the same spatial mode using a polarizing beamsplitter (PBS), generating N00N states in the diagonal polarization basis. The two photons are then sent through the liquid crystal device which alters the relative phase between the two modes according to the applied voltage. A projective measurement is performed by means of a second HWP and PBS. We record the coincidences counts corresponding to the postselected outcome probabilities: 
\begin{equation}
P_{\theta}(\varphi,v(x))=\frac{1}{4}\left(1+v(x)\cos{(8\theta-2\varphi(x))}\right),
\end{equation}
where $v$ is the fringe visibility, setting the HWP to $\theta=\left\lbrace0,\pi/16,\pi/8,3\pi/16\right\rbrace$.
For the individual photon measurements, the $\vert H \rangle$ polarized photon is still sent through the same setup, while the other is directly coupled to the detector for heralding. \\

\textbf{Detection.} The output modes of the interferometer are coupled to single mode fibers and sent to two Avalanche Photodiodes (APDs) for detection. The acquisition time is set to 3 s for the reference measurement and to 0.5 s for the two measurements at lower $N_r$. Coincidences are recorded by means of a Field Programmable Gate Array (FPGA).\\

\textbf{Bayesian estimation.} In order improve the robustness against the setup instabilities, we perform a Bayesian multiparameter estimation routine, obtaining the values for the phase, fringe visibility, and their respective variances at each fiducial point. These are calculated starting from the Bayesian probability defined as:
\begin{equation}
    P_B(\varphi(x),v(x))=\mathcal{N}\prod_{\theta}P_{\theta}(\varphi(x),v(x))^{n_{\theta}}
\end{equation}
where we have considered the use of a uniform {\it a priori} probability distribution, $\mathcal{N}$ is a normalization constant, and $n_{\theta}$ are the measured coincidences for the $\theta$-th projection. The first and second moments of this distribution, yield the desired quantities:
\begin{equation}
\begin{aligned}
&\varphi_B(x)= \int \varphi(x) P_B(\varphi(x),v(x)) d\varphi(x)dv(x)\\
&v_B(x)=\int v(x) P_B(\varphi(x),v(x)) d\varphi(x)dv(x)\\
&\Delta^2\varphi_B(x)= \int (\varphi(x)-\varphi_B(x))^2 P_B(\varphi(x),v(x)) d\varphi(x)dv(x)\\
&\Delta^2v_B(x)=\int (v(x)-v_B(x))^2 P_B(\varphi(x),v(x)) d\varphi(x)dv(x)\\
\end{aligned}
\end{equation}
\\

\textbf{Error evaluation.} The uncertainties on the error $\delta^2$, are obtained performing a Montercarlo routine as follows:500 sets of the estimated phase values are generated by adding a random Gaussian distributed error with variance $\Delta^2\varphi_B(x)$ to the estimated values $\varphi_B(x)$. For each set of phases, the resampling and interpolation procedures are then performed as described in the main text, and the error $\delta^2$ is calculated. The error on $\delta^2$ is hence obtained from the standard deviation over the 500 repetitions of $\delta^2$.
\\

\textbf{Simulations.} Simulations have been carried out to determine the expected values of $\delta^2$ for ideally estimated fiducial points. We have employed the measured $\varphi_{\rm ref}(x)$ both as reference and as set of data from which we have selected $N_s$ sampling points. To simulate experimental data, we have performed a Montecarlo routine adding to the sampled points a random Gaussian distributed error with variance dictated by the Cramér-Rao bound: $\varepsilon_i^2=1/N_rF_i$, where $F_i$ is the Fisher information for the classical and quantum measurements, which reads~\cite{Roccia2018}:
\begin{equation}
    \begin{aligned}
    & F_c=\frac{2v(x)^2}{4-v(x)^2\left(1-\cos{4\varphi(x)}\right)} \\
    & F_q=\frac{8v(x)^2}{4-v(x)^2\left(1-\cos{8\varphi(x)}\right)}.
    \end{aligned}
\end{equation}
We have then performed the interpolation procedure as described in the main text. The obtained simulated mean values are shown as dotted lines in Fig. 3 of the main text.
\\

\bibliography{mainbib}

\begin{thebibliography}{30}%
\makeatletter
\providecommand \@ifxundefined [1]{%
 \@ifx{#1\undefined}
}%
\providecommand \@ifnum [1]{%
 \ifnum #1\expandafter \@firstoftwo
 \else \expandafter \@secondoftwo
 \fi
}%
\providecommand \@ifx [1]{%
 \ifx #1\expandafter \@firstoftwo
 \else \expandafter \@secondoftwo
 \fi
}%
\providecommand \natexlab [1]{#1}%
\providecommand \enquote  [1]{``#1''}%
\providecommand \bibnamefont  [1]{#1}%
\providecommand \bibfnamefont [1]{#1}%
\providecommand \citenamefont [1]{#1}%
\providecommand \href@noop [0]{\@secondoftwo}%
\providecommand \href [0]{\begingroup \@sanitize@url \@href}%
\providecommand \@href[1]{\@@startlink{#1}\@@href}%
\providecommand \@@href[1]{\endgroup#1\@@endlink}%
\providecommand \@sanitize@url [0]{\catcode `\\12\catcode `\$12\catcode
  `\&12\catcode `\#12\catcode `\^12\catcode `\_12\catcode `\%12\relax}%
\providecommand \@@startlink[1]{}%
\providecommand \@@endlink[0]{}%
\providecommand \url  [0]{\begingroup\@sanitize@url \@url }%
\providecommand \@url [1]{\endgroup\@href {#1}{\urlprefix }}%
\providecommand \urlprefix  [0]{URL }%
\providecommand \Eprint [0]{\href }%
\providecommand \doibase [0]{http://dx.doi.org/}%
\providecommand \selectlanguage [0]{\@gobble}%
\providecommand \bibinfo  [0]{\@secondoftwo}%
\providecommand \bibfield  [0]{\@secondoftwo}%
\providecommand \translation [1]{[#1]}%
\providecommand \BibitemOpen [0]{}%
\providecommand \bibitemStop [0]{}%
\providecommand \bibitemNoStop [0]{.\EOS\space}%
\providecommand \EOS [0]{\spacefactor3000\relax}%
\providecommand \BibitemShut  [1]{\csname bibitem#1\endcsname}%
\let\auto@bib@innerbib\@empty
\bibitem [{\citenamefont {Giovannetti}\ \emph {et~al.}(2004)\citenamefont
  {Giovannetti}, \citenamefont {Lloyd},\ and\ \citenamefont
  {Maccone}}]{Giovannetti1}%
  \BibitemOpen
  \bibfield  {author} {\bibinfo {author} {\bibfnamefont {V.}~\bibnamefont
  {Giovannetti}}, \bibinfo {author} {\bibfnamefont {S.}~\bibnamefont {Lloyd}},
  \ and\ \bibinfo {author} {\bibfnamefont {L.}~\bibnamefont {Maccone}},\ }\href
  {\doibase 10.1126/science.1104149} {\bibfield  {journal} {\bibinfo  {journal}
  {Science}\ }\textbf {\bibinfo {volume} {306}},\ \bibinfo {pages} {1330}
  (\bibinfo {year} {2004})}\BibitemShut {NoStop}%
\bibitem [{\citenamefont {Giovannetti}\ \emph {et~al.}(2006)\citenamefont
  {Giovannetti}, \citenamefont {Lloyd},\ and\ \citenamefont
  {Maccone}}]{Giovannetti3}%
  \BibitemOpen
  \bibfield  {author} {\bibinfo {author} {\bibfnamefont {V.}~\bibnamefont
  {Giovannetti}}, \bibinfo {author} {\bibfnamefont {S.}~\bibnamefont {Lloyd}},
  \ and\ \bibinfo {author} {\bibfnamefont {L.}~\bibnamefont {Maccone}},\ }\href
  {\doibase 10.1103/PhysRevLett.96.010401} {\bibfield  {journal} {\bibinfo
  {journal} {Phys. Rev. Lett.}\ }\textbf {\bibinfo {volume} {96}},\ \bibinfo
  {pages} {010401} (\bibinfo {year} {2006})}\BibitemShut {NoStop}%
\bibitem [{\citenamefont {Paris}(2009)}]{paris}%
  \BibitemOpen
  \bibfield  {author} {\bibinfo {author} {\bibfnamefont {M.~G.~A.}\
  \bibnamefont {Paris}},\ }\href {\doibase 10.1142/S0219749909004839}
  {\bibfield  {journal} {\bibinfo  {journal} {Int. J. Quantum Inf.}\ }\textbf
  {\bibinfo {volume} {07}},\ \bibinfo {pages} {125} (\bibinfo {year}
  {2009})}\BibitemShut {NoStop}%
\bibitem [{\citenamefont {Lee}\ \emph {et~al.}(2002)\citenamefont {Lee},
  \citenamefont {Kok},\ and\ \citenamefont {Dowling}}]{lee}%
  \BibitemOpen
  \bibfield  {author} {\bibinfo {author} {\bibfnamefont {H.}~\bibnamefont
  {Lee}}, \bibinfo {author} {\bibfnamefont {P.}~\bibnamefont {Kok}}, \ and\
  \bibinfo {author} {\bibfnamefont {J.~P.}\ \bibnamefont {Dowling}},\ }\href
  {\doibase 10.1080/0950034021000011536} {\bibfield  {journal} {\bibinfo
  {journal} {J. Mod. Opt.}\ }\textbf {\bibinfo {volume} {49}},\ \bibinfo
  {pages} {2325} (\bibinfo {year} {2002})}\BibitemShut {NoStop}%
\bibitem [{\citenamefont {Kura}\ and\ \citenamefont {Ueda}(2020)}]{ueda}%
  \BibitemOpen
  \bibfield  {author} {\bibinfo {author} {\bibfnamefont {N.}~\bibnamefont
  {Kura}}\ and\ \bibinfo {author} {\bibfnamefont {M.}~\bibnamefont {Ueda}},\
  }\href {\doibase 10.1103/PhysRevLett.124.010507} {\bibfield  {journal}
  {\bibinfo  {journal} {Phys. Rev. Lett.}\ }\textbf {\bibinfo {volume} {124}},\
  \bibinfo {pages} {010507} (\bibinfo {year} {2020})}\BibitemShut {NoStop}%
\bibitem [{\citenamefont {Tsang}\ \emph {et~al.}(2011)\citenamefont {Tsang},
  \citenamefont {Wiseman},\ and\ \citenamefont {Caves}}]{Tsang2011}%
  \BibitemOpen
  \bibfield  {author} {\bibinfo {author} {\bibfnamefont {M.}~\bibnamefont
  {Tsang}}, \bibinfo {author} {\bibfnamefont {H.~M.}\ \bibnamefont {Wiseman}},
  \ and\ \bibinfo {author} {\bibfnamefont {C.~M.}\ \bibnamefont {Caves}},\
  }\href {\doibase 10.1103/PhysRevLett.106.090401} {\bibfield  {journal}
  {\bibinfo  {journal} {Phys. Rev. Lett.}\ }\textbf {\bibinfo {volume} {106}},\
  \bibinfo {pages} {090401} (\bibinfo {year} {2011})}\BibitemShut {NoStop}%
\bibitem [{\citenamefont {Polino}\ \emph {et~al.}(2020)\citenamefont {Polino},
  \citenamefont {Valeri}, \citenamefont {Spagnolo},\ and\ \citenamefont
  {Sciarrino}}]{Polinoreview}%
  \BibitemOpen
  \bibfield  {author} {\bibinfo {author} {\bibfnamefont {E.}~\bibnamefont
  {Polino}}, \bibinfo {author} {\bibfnamefont {M.}~\bibnamefont {Valeri}},
  \bibinfo {author} {\bibfnamefont {N.}~\bibnamefont {Spagnolo}}, \ and\
  \bibinfo {author} {\bibfnamefont {F.}~\bibnamefont {Sciarrino}},\ }\href
  {\doibase 10.1116/5.0007577} {\bibfield  {journal} {\bibinfo  {journal} {AVS
  Quantum Sci.}\ }\textbf {\bibinfo {volume} {2}},\ \bibinfo {pages} {024703}
  (\bibinfo {year} {2020})}\BibitemShut {NoStop}%
\bibitem [{\citenamefont {Albarelli}\ \emph {et~al.}(2020)\citenamefont
  {Albarelli}, \citenamefont {Barbieri}, \citenamefont {Genoni},\ and\
  \citenamefont {Gianani}}]{multireview}%
  \BibitemOpen
  \bibfield  {author} {\bibinfo {author} {\bibfnamefont {F.}~\bibnamefont
  {Albarelli}}, \bibinfo {author} {\bibfnamefont {M.}~\bibnamefont {Barbieri}},
  \bibinfo {author} {\bibfnamefont {M.~G.}\ \bibnamefont {Genoni}}, \ and\
  \bibinfo {author} {\bibfnamefont {I.}~\bibnamefont {Gianani}},\ }\href
  {\doibase 10.1016/j.physleta.2020.126311} {\bibfield  {journal} {\bibinfo
  {journal} {Phys. Lett. A}\ }\textbf {\bibinfo {volume} {384}},\ \bibinfo
  {pages} {126311} (\bibinfo {year} {2020})}\BibitemShut {NoStop}%
\bibitem [{\citenamefont {Pirandola}\ \emph {et~al.}(2018)\citenamefont
  {Pirandola}, \citenamefont {Bardhan}, \citenamefont {Gehring}, \citenamefont
  {Weedbrook},\ and\ \citenamefont {Lloyd}}]{pirandola}%
  \BibitemOpen
  \bibfield  {author} {\bibinfo {author} {\bibfnamefont {S.}~\bibnamefont
  {Pirandola}}, \bibinfo {author} {\bibfnamefont {B.~R.}\ \bibnamefont
  {Bardhan}}, \bibinfo {author} {\bibfnamefont {T.}~\bibnamefont {Gehring}},
  \bibinfo {author} {\bibfnamefont {C.}~\bibnamefont {Weedbrook}}, \ and\
  \bibinfo {author} {\bibfnamefont {S.}~\bibnamefont {Lloyd}},\ }\href
  {\doibase 10.1038/s41566-018-0301-6} {\bibfield  {journal} {\bibinfo
  {journal} {Nat. Photonics}\ }\textbf {\bibinfo {volume} {12}},\ \bibinfo
  {pages} {724} (\bibinfo {year} {2018})}\BibitemShut {NoStop}%
\bibitem [{\citenamefont {Demkowicz-Dobrza{\'{n}}ski}\ \emph
  {et~al.}(2015)\citenamefont {Demkowicz-Dobrza{\'{n}}ski}, \citenamefont
  {Jarzyna},\ and\ \citenamefont
  {Ko{\l}ody{\'{n}}ski}}]{Demkowicz-Dobrzanski2015a}%
  \BibitemOpen
  \bibfield  {author} {\bibinfo {author} {\bibfnamefont {R.}~\bibnamefont
  {Demkowicz-Dobrza{\'{n}}ski}}, \bibinfo {author} {\bibfnamefont
  {M.}~\bibnamefont {Jarzyna}}, \ and\ \bibinfo {author} {\bibfnamefont
  {J.}~\bibnamefont {Ko{\l}ody{\'{n}}ski}},\ }in\ \href {\doibase
  10.1016/bs.po.2015.02.003} {\emph {\bibinfo {booktitle} {Prog. Opt. Vol.
  60}}},\ \bibinfo {editor} {edited by\ \bibinfo {editor} {\bibfnamefont
  {E.}~\bibnamefont {Wolf}}}\ (\bibinfo  {publisher} {Elsevier},\ \bibinfo
  {address} {Amsterdam},\ \bibinfo {year} {2015})\ Chap.~\bibinfo {chapter}
  {4}, pp.\ \bibinfo {pages} {345--435}\BibitemShut {NoStop}%
\bibitem [{\citenamefont {Pezz\`e}\ \emph {et~al.}(2018)\citenamefont
  {Pezz\`e}, \citenamefont {Smerzi}, \citenamefont {Oberthaler}, \citenamefont
  {Schmied},\ and\ \citenamefont {Treutlein}}]{atomicensembles}%
  \BibitemOpen
  \bibfield  {author} {\bibinfo {author} {\bibfnamefont {L.}~\bibnamefont
  {Pezz\`e}}, \bibinfo {author} {\bibfnamefont {A.}~\bibnamefont {Smerzi}},
  \bibinfo {author} {\bibfnamefont {M.~K.}\ \bibnamefont {Oberthaler}},
  \bibinfo {author} {\bibfnamefont {R.}~\bibnamefont {Schmied}}, \ and\
  \bibinfo {author} {\bibfnamefont {P.}~\bibnamefont {Treutlein}},\ }\href
  {\doibase 10.1103/RevModPhys.90.035005} {\bibfield  {journal} {\bibinfo
  {journal} {Rev. Mod. Phys.}\ }\textbf {\bibinfo {volume} {90}},\ \bibinfo
  {pages} {035005} (\bibinfo {year} {2018})}\BibitemShut {NoStop}%
\bibitem [{\citenamefont {Degen}\ \emph {et~al.}(2017)\citenamefont {Degen},
  \citenamefont {Reinhard},\ and\ \citenamefont {Cappellaro}}]{qsensingreview}%
  \BibitemOpen
  \bibfield  {author} {\bibinfo {author} {\bibfnamefont {C.~L.}\ \bibnamefont
  {Degen}}, \bibinfo {author} {\bibfnamefont {F.}~\bibnamefont {Reinhard}}, \
  and\ \bibinfo {author} {\bibfnamefont {P.}~\bibnamefont {Cappellaro}},\
  }\href {\doibase 10.1103/RevModPhys.89.035002} {\bibfield  {journal}
  {\bibinfo  {journal} {Rev. Mod. Phys.}\ }\textbf {\bibinfo {volume} {89}},\
  \bibinfo {pages} {035002} (\bibinfo {year} {2017})}\BibitemShut {NoStop}%
\bibitem [{\citenamefont {Humphreys}\ \emph {et~al.}(2013)\citenamefont
  {Humphreys}, \citenamefont {Barbieri}, \citenamefont {Datta},\ and\
  \citenamefont {Walmsley}}]{peter}%
  \BibitemOpen
  \bibfield  {author} {\bibinfo {author} {\bibfnamefont {P.~C.}\ \bibnamefont
  {Humphreys}}, \bibinfo {author} {\bibfnamefont {M.}~\bibnamefont {Barbieri}},
  \bibinfo {author} {\bibfnamefont {A.}~\bibnamefont {Datta}}, \ and\ \bibinfo
  {author} {\bibfnamefont {I.~A.}\ \bibnamefont {Walmsley}},\ }\href {\doibase
  10.1103/PhysRevLett.111.070403} {\bibfield  {journal} {\bibinfo  {journal}
  {Phys. Rev. Lett.}\ }\textbf {\bibinfo {volume} {111}},\ \bibinfo {pages}
  {070403} (\bibinfo {year} {2013})}\BibitemShut {NoStop}%
\bibitem [{\citenamefont {Wheatley}\ \emph {et~al.}(2010)\citenamefont
  {Wheatley}, \citenamefont {Berry}, \citenamefont {Yonezawa}, \citenamefont
  {Nakane}, \citenamefont {Arao}, \citenamefont {Pope}, \citenamefont {Ralph},
  \citenamefont {Wiseman}, \citenamefont {Furusawa},\ and\ \citenamefont
  {Huntington}}]{Wheatley2010}%
  \BibitemOpen
  \bibfield  {author} {\bibinfo {author} {\bibfnamefont {T.~A.}\ \bibnamefont
  {Wheatley}}, \bibinfo {author} {\bibfnamefont {D.~W.}\ \bibnamefont {Berry}},
  \bibinfo {author} {\bibfnamefont {H.}~\bibnamefont {Yonezawa}}, \bibinfo
  {author} {\bibfnamefont {D.}~\bibnamefont {Nakane}}, \bibinfo {author}
  {\bibfnamefont {H.}~\bibnamefont {Arao}}, \bibinfo {author} {\bibfnamefont
  {D.~T.}\ \bibnamefont {Pope}}, \bibinfo {author} {\bibfnamefont {T.~C.}\
  \bibnamefont {Ralph}}, \bibinfo {author} {\bibfnamefont {H.~M.}\ \bibnamefont
  {Wiseman}}, \bibinfo {author} {\bibfnamefont {A.}~\bibnamefont {Furusawa}}, \
  and\ \bibinfo {author} {\bibfnamefont {E.~H.}\ \bibnamefont {Huntington}},\
  }\href {\doibase 10.1103/PhysRevLett.104.093601} {\bibfield  {journal}
  {\bibinfo  {journal} {Phys. Rev. Lett.}\ }\textbf {\bibinfo {volume} {104}},\
  \bibinfo {pages} {093601} (\bibinfo {year} {2010})}\BibitemShut {NoStop}%
\bibitem [{\citenamefont {Yonezawa}\ \emph {et~al.}(2012)\citenamefont
  {Yonezawa}, \citenamefont {Nakane}, \citenamefont {Wheatley}, \citenamefont
  {Iwasawa}, \citenamefont {Takeda}, \citenamefont {Arao}, \citenamefont
  {Ohki}, \citenamefont {Tsumura}, \citenamefont {Berry}, \citenamefont
  {Ralph}, \citenamefont {Wiseman}, \citenamefont {Huntington},\ and\
  \citenamefont {Furusawa}}]{Yonezawa2012}%
  \BibitemOpen
  \bibfield  {author} {\bibinfo {author} {\bibfnamefont {H.}~\bibnamefont
  {Yonezawa}}, \bibinfo {author} {\bibfnamefont {D.}~\bibnamefont {Nakane}},
  \bibinfo {author} {\bibfnamefont {T.~A.}\ \bibnamefont {Wheatley}}, \bibinfo
  {author} {\bibfnamefont {K.}~\bibnamefont {Iwasawa}}, \bibinfo {author}
  {\bibfnamefont {S.}~\bibnamefont {Takeda}}, \bibinfo {author} {\bibfnamefont
  {H.}~\bibnamefont {Arao}}, \bibinfo {author} {\bibfnamefont {K.}~\bibnamefont
  {Ohki}}, \bibinfo {author} {\bibfnamefont {K.}~\bibnamefont {Tsumura}},
  \bibinfo {author} {\bibfnamefont {D.~W.}\ \bibnamefont {Berry}}, \bibinfo
  {author} {\bibfnamefont {T.~C.}\ \bibnamefont {Ralph}}, \bibinfo {author}
  {\bibfnamefont {H.~M.}\ \bibnamefont {Wiseman}}, \bibinfo {author}
  {\bibfnamefont {E.~H.}\ \bibnamefont {Huntington}}, \ and\ \bibinfo {author}
  {\bibfnamefont {A.}~\bibnamefont {Furusawa}},\ }\href {\doibase
  10.1126/science.1225258} {\bibfield  {journal} {\bibinfo  {journal}
  {Science}\ }\textbf {\bibinfo {volume} {337}},\ \bibinfo {pages} {1514}
  (\bibinfo {year} {2012})}\BibitemShut {NoStop}%
\bibitem [{\citenamefont {Cimini}\ \emph {et~al.}(2019)\citenamefont {Cimini},
  \citenamefont {Mellini}, \citenamefont {Rampioni}, \citenamefont {Sbroscia},
  \citenamefont {Leoni}, \citenamefont {Barbieri},\ and\ \citenamefont
  {Gianani}}]{Cimini20191}%
  \BibitemOpen
  \bibfield  {author} {\bibinfo {author} {\bibfnamefont {V.}~\bibnamefont
  {Cimini}}, \bibinfo {author} {\bibfnamefont {M.}~\bibnamefont {Mellini}},
  \bibinfo {author} {\bibfnamefont {G.}~\bibnamefont {Rampioni}}, \bibinfo
  {author} {\bibfnamefont {M.}~\bibnamefont {Sbroscia}}, \bibinfo {author}
  {\bibfnamefont {L.}~\bibnamefont {Leoni}}, \bibinfo {author} {\bibfnamefont
  {M.}~\bibnamefont {Barbieri}}, \ and\ \bibinfo {author} {\bibfnamefont
  {I.}~\bibnamefont {Gianani}},\ }\href {\doibase 10.1364/OE.27.035245}
  {\bibfield  {journal} {\bibinfo  {journal} {Opt. Express}\ }\textbf {\bibinfo
  {volume} {27}},\ \bibinfo {pages} {35245} (\bibinfo {year}
  {2019})}\BibitemShut {NoStop}%
\bibitem [{\citenamefont {Petersen}\ and\ \citenamefont
  {M{\o}lmer}(2006)}]{Petersen2006}%
  \BibitemOpen
  \bibfield  {author} {\bibinfo {author} {\bibfnamefont {V.}~\bibnamefont
  {Petersen}}\ and\ \bibinfo {author} {\bibfnamefont {K.}~\bibnamefont
  {M{\o}lmer}},\ }\href {\doibase 10.1103/PhysRevA.74.043802} {\bibfield
  {journal} {\bibinfo  {journal} {Phys. Rev. A}\ }\textbf {\bibinfo {volume}
  {74}},\ \bibinfo {pages} {043802} (\bibinfo {year} {2006})}\BibitemShut
  {NoStop}%
\bibitem [{\citenamefont {Berry}\ \emph {et~al.}(2013)\citenamefont {Berry},
  \citenamefont {Hall},\ and\ \citenamefont {Wiseman}}]{Berry2013a}%
  \BibitemOpen
  \bibfield  {author} {\bibinfo {author} {\bibfnamefont {D.~W.}\ \bibnamefont
  {Berry}}, \bibinfo {author} {\bibfnamefont {M.~J.~W.}\ \bibnamefont {Hall}},
  \ and\ \bibinfo {author} {\bibfnamefont {H.~M.}\ \bibnamefont {Wiseman}},\
  }\href {\doibase 10.1103/PhysRevLett.111.113601} {\bibfield  {journal}
  {\bibinfo  {journal} {Phys. Rev. Lett.}\ }\textbf {\bibinfo {volume} {111}},\
  \bibinfo {pages} {113601} (\bibinfo {year} {2013})}\BibitemShut {NoStop}%
\bibitem [{\citenamefont {Berry}\ \emph {et~al.}(2015)\citenamefont {Berry},
  \citenamefont {Tsang}, \citenamefont {Hall},\ and\ \citenamefont
  {Wiseman}}]{Berry2015a}%
  \BibitemOpen
  \bibfield  {author} {\bibinfo {author} {\bibfnamefont {D.~W.}\ \bibnamefont
  {Berry}}, \bibinfo {author} {\bibfnamefont {M.}~\bibnamefont {Tsang}},
  \bibinfo {author} {\bibfnamefont {M.~J.~W.}\ \bibnamefont {Hall}}, \ and\
  \bibinfo {author} {\bibfnamefont {H.~M.}\ \bibnamefont {Wiseman}},\ }\href
  {\doibase 10.1103/PhysRevX.5.031018} {\bibfield  {journal} {\bibinfo
  {journal} {Phys. Rev. X}\ }\textbf {\bibinfo {volume} {5}},\ \bibinfo {pages}
  {031018} (\bibinfo {year} {2015})}\BibitemShut {NoStop}%
\bibitem [{\citenamefont {Ng}\ \emph {et~al.}(2016)\citenamefont {Ng},
  \citenamefont {Ang}, \citenamefont {Wheatley}, \citenamefont {Yonezawa},
  \citenamefont {Furusawa}, \citenamefont {Huntington},\ and\ \citenamefont
  {Tsang}}]{Ng2016}%
  \BibitemOpen
  \bibfield  {author} {\bibinfo {author} {\bibfnamefont {S.}~\bibnamefont
  {Ng}}, \bibinfo {author} {\bibfnamefont {S.~Z.}\ \bibnamefont {Ang}},
  \bibinfo {author} {\bibfnamefont {T.~A.}\ \bibnamefont {Wheatley}}, \bibinfo
  {author} {\bibfnamefont {H.}~\bibnamefont {Yonezawa}}, \bibinfo {author}
  {\bibfnamefont {A.}~\bibnamefont {Furusawa}}, \bibinfo {author}
  {\bibfnamefont {E.~H.}\ \bibnamefont {Huntington}}, \ and\ \bibinfo {author}
  {\bibfnamefont {M.}~\bibnamefont {Tsang}},\ }\href {\doibase
  10.1103/PhysRevA.93.042121} {\bibfield  {journal} {\bibinfo  {journal} {Phys.
  Rev. A}\ }\textbf {\bibinfo {volume} {93}},\ \bibinfo {pages} {042121}
  (\bibinfo {year} {2016})}\BibitemShut {NoStop}%
\bibitem [{\citenamefont {{Martin Ciurana}}\ \emph {et~al.}(2017)\citenamefont
  {{Martin Ciurana}}, \citenamefont {Colangelo}, \citenamefont
  {Slodi{\v{c}}ka}, \citenamefont {Sewell},\ and\ \citenamefont
  {Mitchell}}]{MartinCiurana2017}%
  \BibitemOpen
  \bibfield  {author} {\bibinfo {author} {\bibfnamefont {F.}~\bibnamefont
  {{Martin Ciurana}}}, \bibinfo {author} {\bibfnamefont {G.}~\bibnamefont
  {Colangelo}}, \bibinfo {author} {\bibfnamefont {L.}~\bibnamefont
  {Slodi{\v{c}}ka}}, \bibinfo {author} {\bibfnamefont {R.~J.}\ \bibnamefont
  {Sewell}}, \ and\ \bibinfo {author} {\bibfnamefont {M.~W.}\ \bibnamefont
  {Mitchell}},\ }\href {\doibase 10.1103/PhysRevLett.119.043603} {\bibfield
  {journal} {\bibinfo  {journal} {Phys. Rev. Lett.}\ }\textbf {\bibinfo
  {volume} {119}},\ \bibinfo {pages} {043603} (\bibinfo {year}
  {2017})}\BibitemShut {NoStop}%
\bibitem [{\citenamefont {Laverick}\ \emph {et~al.}(2018)\citenamefont
  {Laverick}, \citenamefont {Wiseman}, \citenamefont {Dinani},\ and\
  \citenamefont {Berry}}]{Laverick2018}%
  \BibitemOpen
  \bibfield  {author} {\bibinfo {author} {\bibfnamefont {K.~T.}\ \bibnamefont
  {Laverick}}, \bibinfo {author} {\bibfnamefont {H.~M.}\ \bibnamefont
  {Wiseman}}, \bibinfo {author} {\bibfnamefont {H.~T.}\ \bibnamefont {Dinani}},
  \ and\ \bibinfo {author} {\bibfnamefont {D.~W.}\ \bibnamefont {Berry}},\
  }\href {\doibase 10.1103/PhysRevA.97.042334} {\bibfield  {journal} {\bibinfo
  {journal} {Phys. Rev. A}\ }\textbf {\bibinfo {volume} {97}},\ \bibinfo
  {pages} {042334} (\bibinfo {year} {2018})}\BibitemShut {NoStop}%
\bibitem [{\citenamefont {Slussarenko}\ \emph {et~al.}(2017)\citenamefont
  {Slussarenko}, \citenamefont {Weston}, \citenamefont {Chrzanowski},
  \citenamefont {Shalm}, \citenamefont {Verma}, \citenamefont {Nam},\ and\
  \citenamefont {Pryde}}]{sergei}%
  \BibitemOpen
  \bibfield  {author} {\bibinfo {author} {\bibfnamefont {S.}~\bibnamefont
  {Slussarenko}}, \bibinfo {author} {\bibfnamefont {M.~M.}\ \bibnamefont
  {Weston}}, \bibinfo {author} {\bibfnamefont {H.~M.}\ \bibnamefont
  {Chrzanowski}}, \bibinfo {author} {\bibfnamefont {L.~K.}\ \bibnamefont
  {Shalm}}, \bibinfo {author} {\bibfnamefont {V.~B.}\ \bibnamefont {Verma}},
  \bibinfo {author} {\bibfnamefont {S.~W.}\ \bibnamefont {Nam}}, \ and\
  \bibinfo {author} {\bibfnamefont {G.~J.}\ \bibnamefont {Pryde}},\ }\href
  {\doibase 10.1038/s41566-017-0011-5} {\bibfield  {journal} {\bibinfo
  {journal} {Nature Photonics}\ }\textbf {\bibinfo {volume} {11}},\ \bibinfo
  {pages} {700} (\bibinfo {year} {2017})}\BibitemShut {NoStop}%
\bibitem [{\citenamefont {Ramsay}\ and\ \citenamefont
  {Silverman}(2002)}]{libro}%
  \BibitemOpen
  \bibfield  {author} {\bibinfo {author} {\bibfnamefont {J.}~\bibnamefont
  {Ramsay}}\ and\ \bibinfo {author} {\bibfnamefont {B.}~\bibnamefont
  {Silverman}},\ }\href@noop {} {\emph {\bibinfo {title} {Applied Functional
  Data Analysis}}}\ (\bibinfo  {publisher} {Springer-Verlag New York},\
  \bibinfo {year} {2002})\BibitemShut {NoStop}%
\bibitem [{\citenamefont {Szczykulska}\ \emph {et~al.}(2016)\citenamefont
  {Szczykulska}, \citenamefont {Baumgratz},\ and\ \citenamefont
  {Datta}}]{Szczykulska2016}%
  \BibitemOpen
  \bibfield  {author} {\bibinfo {author} {\bibfnamefont {M.}~\bibnamefont
  {Szczykulska}}, \bibinfo {author} {\bibfnamefont {T.}~\bibnamefont
  {Baumgratz}}, \ and\ \bibinfo {author} {\bibfnamefont {A.}~\bibnamefont
  {Datta}},\ }\href {\doibase 10.1080/23746149.2016.1230476} {\bibfield
  {journal} {\bibinfo  {journal} {Adv. Phys. X}\ }\textbf {\bibinfo {volume}
  {1}},\ \bibinfo {pages} {621} (\bibinfo {year} {2016})}\BibitemShut {NoStop}%
\bibitem [{\citenamefont {Demkowicz-Dobrza{\'{n}}ski}\ \emph
  {et~al.}(2020)\citenamefont {Demkowicz-Dobrza{\'{n}}ski}, \citenamefont
  {G{\'{o}}recki},\ and\ \citenamefont {Guta}}]{Demkowicz-Dobrzanski2020}%
  \BibitemOpen
  \bibfield  {author} {\bibinfo {author} {\bibfnamefont {R.}~\bibnamefont
  {Demkowicz-Dobrza{\'{n}}ski}}, \bibinfo {author} {\bibfnamefont
  {W.}~\bibnamefont {G{\'{o}}recki}}, \ and\ \bibinfo {author} {\bibfnamefont
  {M.}~\bibnamefont {Guta}},\ }\href {\doibase 10.1088/1751-8121/ab8ef3}
  {\bibfield  {journal} {\bibinfo  {journal} {J. Phys. A}\ ,\ \bibinfo {pages}
  {in press}} (\bibinfo {year} {2020})}\BibitemShut {NoStop}%
\bibitem [{\citenamefont {Roccia}\ \emph {et~al.}(2018)\citenamefont {Roccia},
  \citenamefont {Cimini}, \citenamefont {Sbroscia}, \citenamefont {Gianani},
  \citenamefont {Ruggiero}, \citenamefont {Mancino}, \citenamefont {Genoni},
  \citenamefont {Ricci},\ and\ \citenamefont {Barbieri}}]{Roccia2018}%
  \BibitemOpen
  \bibfield  {author} {\bibinfo {author} {\bibfnamefont {E.}~\bibnamefont
  {Roccia}}, \bibinfo {author} {\bibfnamefont {V.}~\bibnamefont {Cimini}},
  \bibinfo {author} {\bibfnamefont {M.}~\bibnamefont {Sbroscia}}, \bibinfo
  {author} {\bibfnamefont {I.}~\bibnamefont {Gianani}}, \bibinfo {author}
  {\bibfnamefont {L.}~\bibnamefont {Ruggiero}}, \bibinfo {author}
  {\bibfnamefont {L.}~\bibnamefont {Mancino}}, \bibinfo {author} {\bibfnamefont
  {M.~G.}\ \bibnamefont {Genoni}}, \bibinfo {author} {\bibfnamefont {M.~A.}\
  \bibnamefont {Ricci}}, \ and\ \bibinfo {author} {\bibfnamefont
  {M.}~\bibnamefont {Barbieri}},\ }\href {\doibase 10.1364/OPTICA.5.001171}
  {\bibfield  {journal} {\bibinfo  {journal} {Optica}\ }\textbf {\bibinfo
  {volume} {5}},\ \bibinfo {pages} {1171} (\bibinfo {year} {2018})}\BibitemShut
  {NoStop}%
\bibitem [{\citenamefont {Tsang}\ \emph {et~al.}(2019)\citenamefont {Tsang},
  \citenamefont {Albarelli},\ and\ \citenamefont {Datta}}]{Tsang2019}%
  \BibitemOpen
  \bibfield  {author} {\bibinfo {author} {\bibfnamefont {M.}~\bibnamefont
  {Tsang}}, \bibinfo {author} {\bibfnamefont {F.}~\bibnamefont {Albarelli}}, \
  and\ \bibinfo {author} {\bibfnamefont {A.}~\bibnamefont {Datta}},\ }\href
  {http://arxiv.org/abs/1906.09871} {\bibfield  {journal} {\bibinfo  {journal}
  {Phys. Rev. X}\ ,\ \bibinfo {pages} {in press}} (\bibinfo {year}
  {2019})}\BibitemShut {NoStop}%
\bibitem [{\citenamefont {Tsang}(2019)}]{Tsang2019b}%
  \BibitemOpen
  \bibfield  {author} {\bibinfo {author} {\bibfnamefont {M.}~\bibnamefont
  {Tsang}},\ }\href {\doibase 10.1103/PhysRevResearch.1.033006} {\bibfield
  {journal} {\bibinfo  {journal} {Phys. Rev. Res.}\ }\textbf {\bibinfo {volume}
  {1}},\ \bibinfo {pages} {033006} (\bibinfo {year} {2019})}\BibitemShut
  {NoStop}%
\bibitem [{\citenamefont {Taylor}\ and\ \citenamefont
  {Bowen}(2016)}]{Taylor2016}%
  \BibitemOpen
  \bibfield  {author} {\bibinfo {author} {\bibfnamefont {M.~A.}\ \bibnamefont
  {Taylor}}\ and\ \bibinfo {author} {\bibfnamefont {W.~P.}\ \bibnamefont
  {Bowen}},\ }\href {\doibase 10.1016/j.physrep.2015.12.002} {\bibfield
  {journal} {\bibinfo  {journal} {Phys. Rep.}\ }\textbf {\bibinfo {volume}
  {615}},\ \bibinfo {pages} {1} (\bibinfo {year} {2016})}\BibitemShut {NoStop}%
\end{thebibliography}%

\section*{Acknowledgements}
The authors thank Fabio Sciarrino for the loan of scientific equipment, Adriano Verna and Rafal Demkowicz-Dobrzanski for valuable discussion, and M. De Seta, L. Di Gaspare and I. Agha for useful comments.

I.G. is supported by Ministero dell’Istruzione, dell’Università e della Ricerca Grant of Excellence Departments (ARTICOLO 1, COMMI 314-337 LEGGE 232/2016).
F.A. acknowledges financial support from the National Science Center (Poland) grant No. 2016/22/E/ST2/00559.

\section*{Author contributions}

I.G., F.A. and M.B conceived the project. I.G. performed the experiment with assistance from V.C. and M.B. All authors discussed the results and contributed to writing the manuscript.

\section*{Additional Information}
The authors declare no competing financial interests. Correspondence and requests for materials should be addressed to I.G. or M.B.

\end{document}